# Digital ultrasonically encoded (DUE) optical focusing into random media

Jian Wei Tay†, Puxiang Lai†, Yuta Suzuki & Lihong V. Wang*

*Optical Imaging Laboratory, Department of Biomedical Engineering, Washington University in St. Louis, St. Louis, Missouri 63130-4899*

† Equal contribution

*Corresponding author: lhwang@wustl.edu

**Focusing light into opaque random or scattering media such as biological tissue is a much sought-after goal for biomedical applications such as photodynamic therapy[1-3], optical manipulation[4], and photostimulation[5]. However, focusing with conventional lenses is restricted to one transport mean free path[6] in scattering media, limiting both optical penetration depth and resolution. Focusing deeper is possible by using optical phase conjugation[7-12] or wavefront shaping[13-15] to compensate for the scattering. For practical applications, wavefront shaping offers the advantage of a robust optical system that is less sensitive to optical misalignment. Here, the phase of the incident light is spatially tailored using a phase-shifting array to pre-compensate for scattering. The challenge, then, is to determine the phase pattern which allows light to be optimally delivered to the target region. Optimization algorithms are typically employed for this purpose, with visible particles used as targets to generate feedback. However, using these particles is invasive, and light delivery is limited to fixed points. Here, we demonstrate a method for non-invasive and dynamic focusing, by using ultrasound encoding as a virtual guide star[7] for feedback to an optimization algorithm[16]. The light intensity at the acoustic focus was**

**increased by an order of magnitude. This technique has broad biomedical applications, such as in optogenetics or photoactivation of drugs.**

One of the greatest challenges in biomedical optics is focusing light to a target region in biological tissue, which mostly falls into the category of random or scattering media. Focusing is important in nearly all disciplines of biomedical optics. In optical therapy and manipulation, having a sufficient photon density in a targeted region determines both efficacy and specificity, while for imaging, the optical spot size determines the resolution. In clear media, such as air, optical lenses focus light by introducing position-dependent phase shifts, which cause the field to add up in phase at the focal point (Fig. 1a). Biological tissue on the other hand, is typically highly scattering, appearing opaque at visible wavelengths. Within such scattering media, the optical wavefront is distorted due to refractive index variations, leading to light diffusion and loss of focus[7] (Fig. 1b). Here, we propose that focusing within the medium can still be achieved by spatially tailoring the phase distribution of the beam using phase-shifting arrays, such as liquid crystal spatial light modulators (SLMs) or digital mirror displays (DMDs), to maximize the feedback from a virtual guide star (Fig. 1c).

Previously, focusing within scattering media was demonstrated using optical phase conjugation[7-12] (OPC) and wavefront shaping[13-15] (WFS) techniques. In OPC, a hologram formed by the light scattered from the medium records the phase distortion due to propagation through the medium. Reading out the hologram generates a backward-propagating phase-conjugated beam that reverses the scattering. However, strict optical alignment is required since the readout beam has to travel back along the original beam path. Furthermore, in digital OPC, hologram recording and readout use separate devices (typically a camera and an SLM), which need to be perfectly aligned to each other[10].

Optical alignment is comparatively much less important in WFS techniques. Here, the incident wavefront is shaped directly to pre-compensate for scattering. The challenge, then, is to determine an optimal spatial phase pattern. In the past, sequential optimization algorithms were used to determine the phase pattern one element at a time[13-15], using the intensity emitted from embedded target particles as feedback. However, these algorithms tend to be susceptible to measurement noise as the signal originates from a small portion of the incident beam[16]. Furthermore, the use of target particles is invasive, and light delivery is restricted to the embedded particle positions.

Here we present a digital ultrasonically encoded (DUE) optical focusing technique which eliminates the need for invasively embedding target particles, by using focused ultrasound as a virtual guide star. Light travelling through the ultrasonic focal zone is frequency-shifted by the acousto-optic effect [17,18], and is used as feedback to an optimization algorithm. Dynamic focusing is possible by simply moving the ultrasonic focus, either mechanically or electronically. Furthermore, in biological tissue, ultrasound has orders of magnitude weaker absorption and scattering[19,20] than light, giving this technique the potential to work at greater depth. We envision that DUE optical focusing could be used along with ultrasound imaging or ultrasound-modulated optical tomography[21] (UOT) to first locate a point of interest and then to optimize light delivery to that region.

The optimal phase pattern is generated using a genetic optimization algorithm[16]. Genetic algorithms mimic natural evolution processes; possible solutions are generated from a population of phase patterns, in which those patterns that result in increased signals are more likely to be used to create new solutions in subsequent iterations. Compared to sequential algorithms, genetic algorithms optimize the phase pattern in parallel. This parallelism results in a faster initial increase, yielding near optimization—such as 90% of the maximum—sooner. These algorithms are also less sensitive to measurement errors as the signal now originates from the whole pattern, rather than a single element.

Our system is shown schematically in Fig. 1d. We used a computer-controlled SLM, divided into 20×20 independently controlled segments, to shape the incident wavefront. The SLM response was calibrated[22] to provide a linear phase shift of $2\pi$ over 191 grayscale values for each segment. As a proof-of-principle demonstration, we used two diffusers to scatter the light. Viewed head on, the diffusers appeared opaque, and no focus was formed after the first diffuser. A clear gelatin sample was inserted between the diffusers for acoustic coupling, and a 6 MHz ultrasonic transducer was used to modulate the light. After the second diffuser, the ultrasonically encoded (UE) light intensity, averaged over 75 acquisitions, was measured using photorefractive (PR) detection[21] (details in the Supplementary Materials), and used as the feedback to the optimization algorithm. By maximizing the UE signal, we maximize the light intensity within the acoustic focal region, thereby forming an optical focal spot in the sample.

Fig. 2 shows the results from the optimization procedure. After 600 iterations, the UE signal was increased by 11 times over the initial randomized value (Fig. 2a). For each iteration, the highest UE signal value was recorded and plotted in Fig. 2b. To visualize the focal spot, we embedded a bar (1 mm in both *x* and *z* dimensions) containing fluorescent quantum dots along the *y*-axis, within the acoustic focal zone. A CCD was used to image the bar, and the resulting images are shown in Fig. 3 along with the corresponding phase maps. The optimized focal spot (Fig. 3f) appears to be larger than the 300 µm transverse width of the acoustic beam, which is likely due to the acoustic sidebands (see Suppl. Fig. 2). The cross-sectional fluorescence intensity when displaying the optimized pattern was also an order of magnitude greater than when displaying a randomized pattern (Fig. 3g). We also found that the fluorescence intensity from the randomized pattern was similar to the intensity from a uniform phase pattern, indicating that the wavefront was completely scrambled by the first diffuser. A video of the evolution of the optimization procedure is provided in the Supplementary Materials.

It was previously shown[11] that the expected increase in intensity $\eta$ is equal to the ratio of the number of independent SLM segments $N$ to the number of speckle grains in the ultrasound focus $M$,

$$\eta = \frac{N+1}{M}. \tag{1}$$

Based on an illumination diameter of 1 mm and a distance of 20 cm from the first diffuser to the acoustic focus, we expected 130 µm speckle grains at the acoustic focal plane. We used 5 pulses at 6 MHz, equivalent to a length of 1.25 mm, to modulate the light. This pulse length, along with the acoustic transverse width, yielded a calculated 22 ultrasonically encoded speckles. Hence, using equation (1), we expected $\eta$ = 18.

From equation (1), we see that increasing the number of blocks on the SLM pattern increases the amount of light that can be delivered to the target region. However, by simulating the system, we found that the number of iterations required to optimize the phase pattern also scaled linearly with the number of controlled blocks (for details see Supplementary Materials). Hence, for a given number of iterations, the intensity increase is the same. We chose to use 20x20 SLM segments for practical reasons.

As with any optimization algorithm, the sample decorrelation time is an important consideration. The algorithm currently takes several hours to terminate, due mainly to the slow PR detection time (~1.8 s per measurement), as well as the time needed for data acquisition and processing (~1.5 s). Therefore, we were restricted to mechanically stable diffusers for this proof of principle demonstration. We note that the algorithm could have been stopped at about 370 iterations, just over half the total iterations, since the signal shows a slow final convergence. In future work, the algorithm speed could be improved by using faster detection methods, such as spectral hole burning[23-25], confocal Fabry-Perot interferometry[26-27], or photoacoustic tomography[28], as well as by using faster spatial light modulators, and data acquisition devices. Such refinements could allow DUE optical focusing to be used in a variety of applications, such as in phototherapy, photoactivation of medicine, and optogenetics.

## Acknowledgements

This work was sponsored in part by the National Academies Keck Futures Initiative grant IS 13 and National Institute of Health grants DP1 EB016986 (NIH Director's Pioneer Award).

## Author Contributions

J.W.T., P.L., and Y.S. designed the experiment. J.W.T. wrote code for the experiment and simulations. P.L. implemented the PR detection system. J.W.T. and P.L. ran the experiment. J.W.T. and Y.S. prepared the manuscript. L.V.W. provided overall supervision. All authors were involved in analysis of the results and manuscript revision.

## Competing financial statement

J.W.T., P. L. and Y.S. declare no competing financial interests. L.V.W. has financial interests in Microphotoacoustics, Inc. and Endra, Inc., which, however, did not support this work.

## Additional information

Supplementary information is available in the online version of the paper. Correspondence and requests for material should be addressed to L.V.W.

# References

1. Dolmans, D. E., Fukumura, D., & Jain, R. K. Photodynamic therapy for cancer. *Nat. Rev. Cancer*, **3,** 380-387 (2003)

2. Zhuo, S., Chen, J., Wu, G., Zhu, X., Jiang, X., & Xie, S. Label-free multiphoton imaging and photoablation of preinvasive cancer cells, *Appl. Phys. Lett.*, **100,** 023703 (2012).

3. Birchler, M., Viti, F., Zardi, L., Spiess, B., & Neri, D. Selective targeting and photocoagulation of ocular angiogenesis mediated by a phage-derived human antibody fragment, *Nat. Biotechnol.*, **17**, 984-988 (1999).

4. Cižmár, T and Dholakia, K. Shaping the light transmission through a multimode optical fibre: complex transformation analysis and applications in biophotonics, *Opt. Express* **19,** 18871-18884 (2011).

5. Williams, J. C. & Denison, T. From optogenetic technologies to neuromodulation therapies, *Sci. Transl. Med.,* **5,** 177ps6 (2013).

6. Wang, L. V., & Wu, H.-I. *Biomedical Optics: Principles and Imaging* (Wiley-Interscience, New Jersey 2007).

7. Xu, X., Liu, H., & Wang, L. V. Time-reversed ultrasonically encoded optical focusing into scattering media, *Nature Photon.*, **5,** 154–157 (2011).

8. Vellekoop, I. M., Cui, M., & Yang, C. Digital optical phase conjugation of fluorescence in turbid tissue, *Appl. Phys. Lett.*, **101,** 81108 (2012).

9. Hsieh, C., Pu, Y., Grange, R., & Psaltis, D. Digital phase conjugation of second harmonic radiation emitted by nanoparticles in turbid media, *Opt Express*, **18,** 533–537 (2010).

10. Judkewitz, B., Wang, Y. M., Horstmeyer, R., Mathy, A., & Yang, C. Speckle-scale focusing in the diffusive regime with time reversal of variance-encoded light (TROVE), *Nature Photon.*, **7,** 300-305 (2013).

11. Wang, Y. M., Judkewitz, B., Dimarzio, C. A., & Yang, C. Deep-tissue focal fluorescence imaging with digitally time-reversed ultrasound-encoded light, *Nat. Commun.*, **3,** 928 (2012).

12. Lai, P., Suzuki, Y., Xu, X., & Wang, L. V. Focused fluorescence excitation with time-reversed ultrasonically encoded light and imaging in thick scattering media, *Laser Phys. Lett.* **10,** 075604 (2013).

13. Vellekoop, I. M., van Putten, E. G, Lagendijk, A., & Mosk, A. P. Demixing light paths inside disordered metamaterials, *Opt. Express*, **16,** 67–80 (2008).

14. Aulbach, J., Gjonaj, B., Johnson, P., & Lagendijk, A. Spatiotemporal focusing in opaque scattering media by wave front shaping with nonlinear feedback, *Opt. Express*, **20,** 29237–29251 (2012).

15. van Putten, E. G., Lagendijk, A., & Mosk, A. P. Optimal concentration of light in turbid materials, *JOSA B*, **28,** 1200–1203 (2011).

16. Conkey, D. B., Brown, A. N., Caravaca-Aguirre, A. M., & Piestun, R. Genetic algorithm optimization for focusing through turbid media in noisy environments, *Opt. Express*, **20,** 4840–4849 (2012).

17. Leutz, W. & Maret, G. Ultrasonic modulation of multiply scattered light, *Physica B*, **204,** 14–19 (1995).

18. Wang, L. V. Mechanisms of ultrasound modulation of multiply scattered coherent light: An analytic model. *Phys. Rev. Lett.* **87,** 043903 (2001).

19. Yao, G., & Wang, L. V. Theoretical and experimental studies of ultrasound-modulated optical tomography in biological tissue, *Appl. Opt.*, **39,** 659–664 (2000).

20. Prince, J. L., & Links, J. M. *Medical Imaging: Systems and Signals* (Pearson Prentice Hall, New Jersey 2006).

21. Lai, P., Xu, X., & Wang, L. V. Ultrasound-modulated optical tomography at new depth. J. Biomed. Opt. **17,** 066006 (2012).

22. Li, Y., Zhang, H., Kim, C., Wagner, K. H., Hemmer, P., & Wang, L. V. Pulsed ultrasound-modulated optical tomography using spectral-hole burning as a narrowband spectral filter, *Appl. Phys. Lett.*, **93,** 11111 (2008).

23. Tay, J. W., Ledingham, P. M., & Longdell, J. J. Coherent optical ultrasound detection with rare-earth ion dopants, Appl. Opt., **49,** 4331–4334 (2010).

24. Zhang, H., Sabooni, M., Rippe, L., Kim. C., Kröll, S., Wang, L. V., & Hemmer, P. R. Slow light for deep tissue imaging with ultrasound modulation, *Appl. Phys. Lett.*, **100,** 131102 (2012).

25. Sakadžić, S., & Wang, L. V. High-resolution ultrasound-modulated optical tomography in biological tissues, *Opt. Lett.*, **29,** 2770–2772 (2004).

26. Rousseau, G., Blouin, A., & Monchalin J.-P. Ultrasound-modulated optical imaging using a high-power pulsed laser and a double-pass confocal Fabry-Perot interferometer, Opt. Lett., **34,** 3445–3447 (2009).

27. Wang, L. V., & Hu, S. Photoacoustic tomography: In vivo imaging from organelles to organs. *Science*, **335**, 1458-1462 (2012)

# Figures:

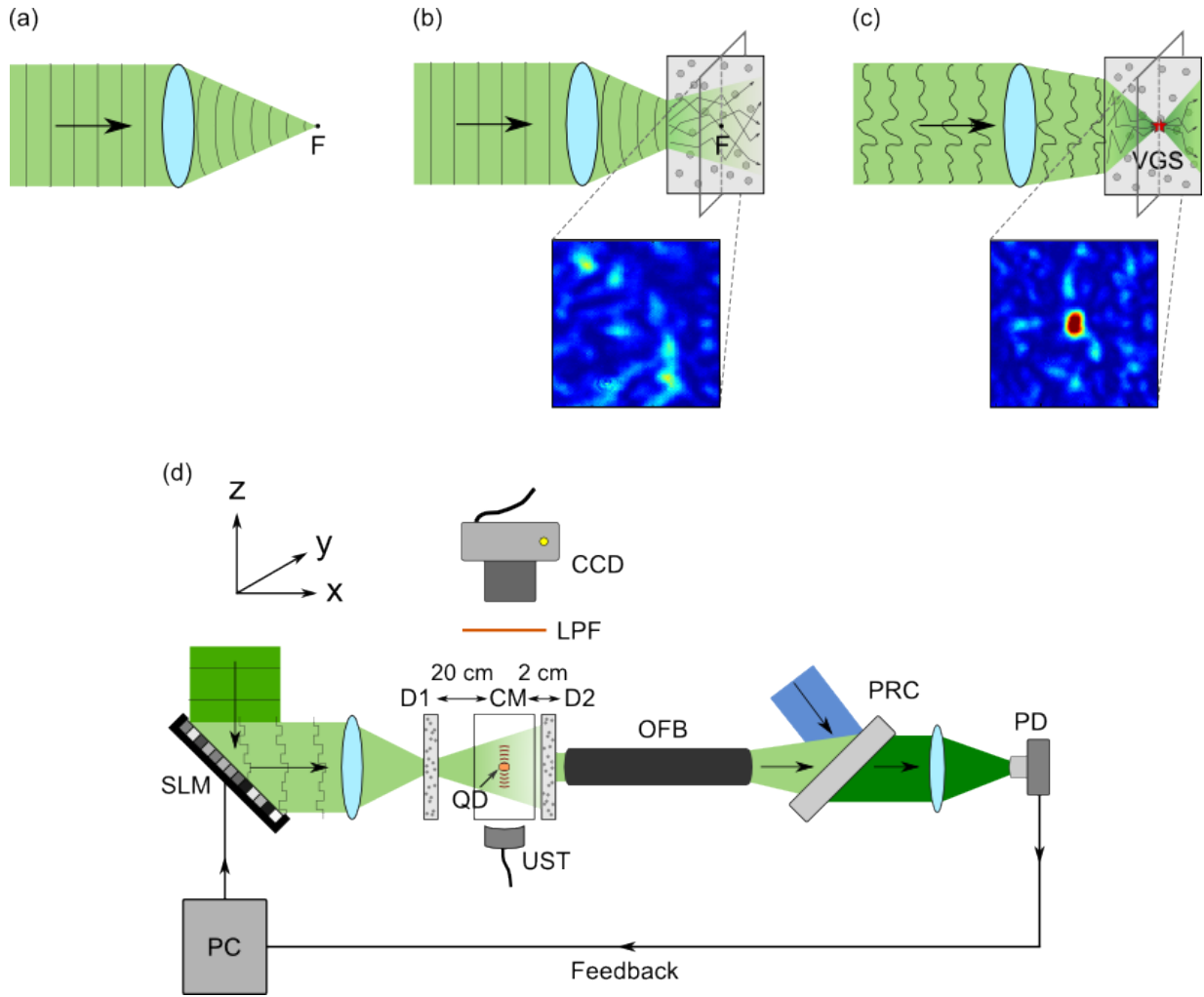

Figure 1: (a) In clear media, an optical lens introduces position-dependent phase shifts which cause the beam to add up constructively at a focal point F. (b) In turbid media, multiple scattering scrambles the wavefront and no focus forms. (c) By maximizing feedback from a virtual guide star (VGS, red star), the incident beam can be spatially tailored to overcome scattering and form a focus within the medium. (d) Schematic of our experiment. The beam incident on the first diffuser (D1) is spatially tailored using a phase-only spatial light modulator (SLM). Five pulses from a 6 MHz ultrasonic transducer (UST) are sent through a clear gelatin medium (CM), modulating light within the acoustic focus. After the second diffuser (D2), the modulated beam is collected using an optical fiber bundle (OFB). The signal is measured using a photorefractive detection (PRC) system and a photodiode (PD), and its amplitude is subsequently used as feedback to optimize the pattern on the SLM. To visualize the focal spot, a bar of fluorescent quantum dots (QD) is embedded along the *y*-axis of the acoustic focal zone. The resulting fluorescence intensity is longpass filtered (LPF) and measured using a CCD.

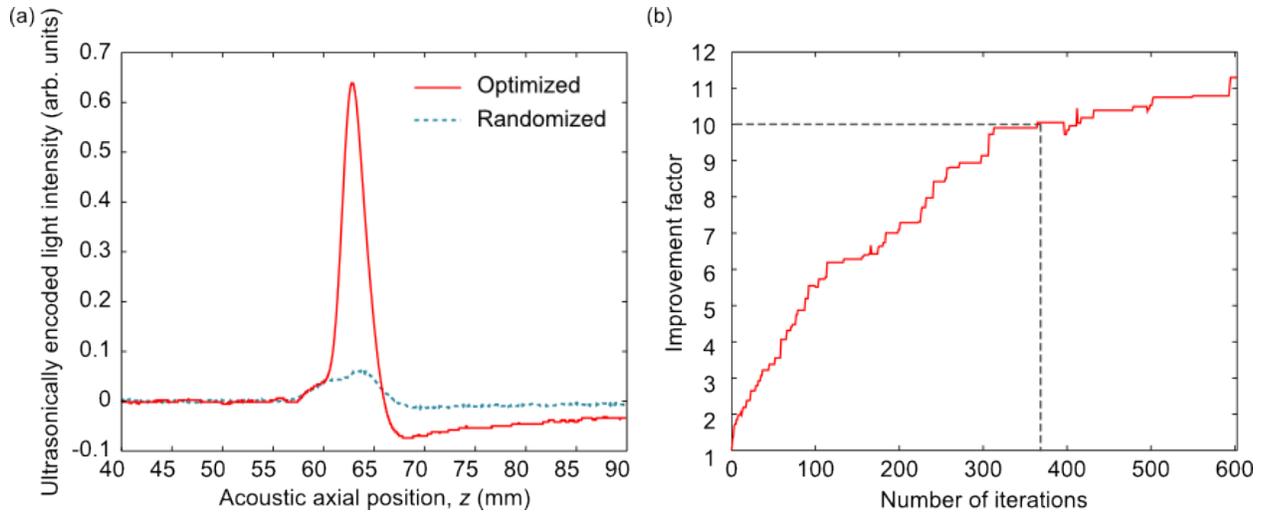

**Figure 2: (a) Measured ultrasonically encoded signal amplitude with the optimized pattern (red, solid line), and with an initial, randomized pattern (blue, dashed line) displayed on the SLM. The traces were averaged over 75 acquisitions. (b) Improvement compared to the initial signal amplitude as the algorithm progresses. An improvement of 11 times in signal amplitude was obtained, which corresponds to a similar increase in light intensity within the acoustic focal zone. An order of magnitude improvement was achieved at 370 iterations.**

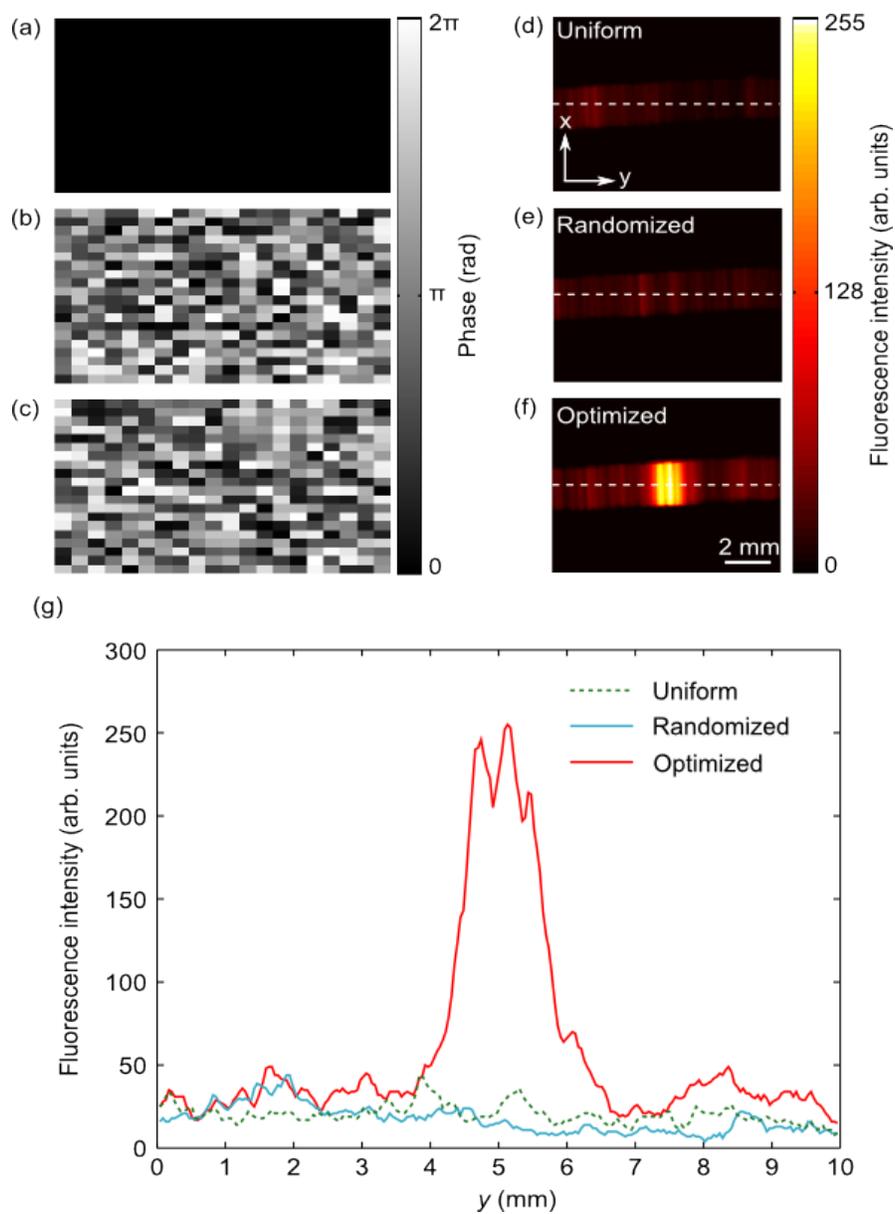

**Figure 3: Visualization of the optimized focal spot.** The phase patterns shown are (a) uniform, (b) randomized, and (c) optimized. Corresponding CCD images of the fluorescent bar are shown in (d-f). Colorbars indicate phase for (a-c), and CCD intensity for (d-f). (g) Cross sectional intensity as indicated by the white dotted lines. An increase of an order of magnitude is seen using the optimized pattern, compared to both uniform and randomized patterns.